\documentclass[12pt]{article}
\def\tbar {\overline{t}}
\def\beq {\begin{equation}}
\def\eeq {\end{equation}}
\def\be {\begin{equation}}
\def\ee {\end{equation}}
\def\barr{\begin{array}}
\def\earr{\end{array}}
\def\bea{\begin{eqnarray}}
\def\eea{\end{eqnarray}}
\def\bmath{\begin{displaymath}}
\def\emath{\end{displaymath}}
\def\bq{\begin{quote}}
\def\eq{\end{quote}}
\def\oas{$O(\alpha_s)$}

\def\g5{\gamma_5}
\def\as{\alpha_s}
\def\real{\mathop{\mbox{\rm Re}}\nolimits}
\def\imag{\mathop{\mbox{\rm Im}}\nolimits}

\def\CF{C_{\scriptscriptstyle F}}

\def\c2t{\cos^2\kern-2pt\theta}

\def\s2t{\sin^2\kern-2pt\theta}

\def\Li{\mbox{$\mbox{\rm Li}_2$}}
\def\Frac#1#2{\mbox{$\textstyle{#1\over#2}$}}
\def\half{\Frac{1}{2}}

\def\nn{\nonumber\\}

\def\MZ{M_{\scriptscriptstyle Z}}

\def\I#1{{I}_{#1}}
\def\S#1{{S}_{#1}}

\def\II#1{{\tilde{I}}_{#1}}
\def\SS#1{{\tilde{S}}_{#1}}

\def\sxi{\mbox{$\sqrt\xi$}}
\def\Deltab#1{{\bf\bar{\rm\Delta}}^{#1}}

\def\Dsigmam{\Delta\sigma^{(-)}}

\begin{document}
\vspace*{-2.5cm}
\begin{flushright}
FTUV-97/8 \\ [-.2cm]
IFIC-97/8 \\ [-.2cm]
PRL-TH-97/6 \\ [-.2cm]
ITP-SB-96-73
\end{flushright}
\vskip.75cm
\begin{center}
{\Large \bf Single-quark spin asymmetries in
            {\boldmath $e^+e^- \rightarrow t\overline{t}$} \\[.25cm]
             and anomalous gluon couplings}
\vskip1cm
Saurabh D.~Rindani$^a$ and Michael M.~Tung$^b$
\vskip.4cm
{\it Instituto de F\'\i sica Corpuscular, Departament de F\'\i sica Te\`orica\\
Universitat de Val\`encia, 46100 Burjassot (Val\`encia), Spain}\\
\vskip.25cm
\it{$^a$Theory Group, Physical Research Laboratory\\
Navrangpura, Ahmedabad 380 009, India}
\vskip.25cm
\it{$^b$Institute for Theoretical Physics\\
State University of New York\\
Stony Brook, NY 11794-3840, U.S.A}
\vskip2cm
{\bf Abstract}
\end{center}
\vskip.25cm
The effect of anomalous chromoelectric couplings of the
gluon to the top quark are considered in $e^+e^- \rightarrow t\overline{t}$.
The total cross section, as well as $t$ and $\tbar$ polarizations are
calculated to order $\alpha_s$ in the presence of the anomalous couplings.
One of the two linear combinations of $t$ and $\tbar$ polarizations is CP
even, while the other is CP odd.
Limits that could be obtained at a future linear collider on 
CP-odd combinations of anomalous couplings are determined.

\newpage
The discovery of a heavy top quark, with a mass of $m_t = 175 \pm 6$ GeV
\cite{CDF}, which is far larger than that of all other quarks, opens up the
possibility that the top quark may have properties very different from those of
the other quarks. Observation of these properties might even signal new physics
beyond the standard model. Several efforts in the past few years have gone
into the investigation of the potential of different experiments to study
possible new interactions of the top quark. In particular, possible anomalous
couplings of the top quark to electroweak gauge bosons\footnote{References
to the voluminous literature on this subject can be found, for example, in
\cite{ttbargamma}.} and
to gluons \cite{ttbarg, rizzog} have also been discussed.  Top polarization is
especially useful in such studies, because with a mass around 175 GeV, the
top quark decays before it can hadronize \cite{bigi}, and all spin information
is preserved in the decay distributions.

In this paper, we investigate the potential of $e^+e^-$ experiments at a future
linear collider with centre-of-mass (c.m.) energies of 500~GeV or higher, to
study the CP-violating anomalous chromoelectric dipole couplings of the top
quark to gluons. So far, a considerable amount of earlier work on the topic of
anomalous gluon couplings has concentrated on hadron colliders. But also
high energy $e^+e^-$ experiments with sufficiently high luminosities would
provide a relatively clean environment to probe the standard model for
anomalous gluon couplings. While earlier efforts in the context of $e^+e^-$
colliders are mainly based on an analysis of the gluon distribution
\cite{rizzog} in $e^+e^-\rightarrow t\tbar g$, we look at the possible
information that could be obtained from studying the total cross section,
and the polarization of $t$ and $\tbar$  separately. This has the advantage
over $t$ and $\tbar$ spin correlations (as for example studied in \cite{parke})
that, because the polarization of only
one of $t$ and $\tbar$ is analyzed by means of a definite decay channel,
the other is free to decay into any channel. This leads to much better
statistics compared to the case when $t-\tbar$ spin correlations are
considered, where definite $t$ and $\tbar$ channels have to be used as
analyzers.

We find that of the three independent quantities, viz., the cross section,
and the $t$ and $\tbar$
polarizations, one quantity, viz., a linear combination of the $t$ and $\tbar$ 
polarizations can be used to probe 
CP-odd chromoelectric dipole coupling. 

We have also considered the effect of beam polarization on the sensitivity
of the measurements.

An effective $t\tbar g$ interaction can be written in the form
\beq
{\cal L}_{t\tbar g} = - g_s \left[ \tbar \gamma^{\mu}G_{\mu}t + \frac{\mu}
{2m_t}\tbar \sigma^{\mu\nu}G_{\mu\nu}t
+ \frac{id}{2m_t}\tbar \sigma^{\mu\nu}\gamma_5 G_{\mu\nu}t \right],
\label{effL}
\eeq
where
\beq
G_{\mu} = \sum_a G^a_{\mu} T_a; \; T^a = \Frac{1}{2} \lambda^a; \;
G_{\mu\nu} = G^a_{\mu\nu}; \; G^a_{\mu\nu} = \partial_{\mu}G^a_{\nu} -
\partial_{\nu}G^a_{\mu};
\eeq
$G^a_{\mu}$ being the gluon field, and $\lambda^a$ being the $SU(3)$ Gell-Mann
matrices. This is the most general Lorentz- and colour-invariant trilinear
interaction (additional quadrilinear terms are needed for local
colour invariance, but we do not need them here). The $\mu$ and $d$ terms
are the chromomagnetic and chromoelectric dipole terms, respectively.
In an effective theory, these are in fact momentum-dependent form factors,
and complex in general.

We use Eq.~({\ref{effL}) to calculate the $t\tbar$ total cross section and the
$t$ and $\tbar$ polarizations  to order $\alpha_s\equiv g_s^2/(4\pi)$.
The extra diagrams contributing to this order are shown in Fig.~1, where the
large dots represent anomalous couplings. The anomalous couplings enter in the
amplitudes for soft and collinear gluon emissions. The infrared divergences in
the amplitudes for soft gluon emission and in the virtual gluon corrections
cancel as in standard QCD, since the anomalous terms vanish in the infrared
limit. Moreover, we do not include anomalous couplings in the loop diagrams.
The latter are included merely to regulate the infrared divergences. We thus
study how anomalous couplings at tree-level would modify standard QCD
predictions at order $\alpha_s$.

For heavy-quark production, the standard QCD one-loop corrections to the total
cross section and the longitudinal spin polarization~\cite{kpt} were calculated
in closed analytic form before. Those results have been extended here to
the case when anomalous couplings are present.

The total unpolarized production rate is given only in terms of the $VV$ and
$AA$ parity-parity combinations for the Born contributions:
\be
\sigma_{Born}\left(e^+ e^-\to\gamma,Z\to q\bar{q}\right) =
\Frac{1}{2}v(3-v^2)\sigma^{VV}+v^3\sigma^{AA},
\ee
where the mass parameters are $v=\sqrt{1-\xi\,}$ and $\xi=4m_q^2/s$.
The $O(\as)$ unpolarized case has the cross section
\be
\sigma\left(e^+ e^-\to\gamma,Z\to q\bar{q}\right) =
\Frac{1}{2}v(3-v^2)\sigma^{VV}c^{VV}+v^3\sigma^{AA}c^{AA},
\ee
where the $VV$ and $AA$ factors that
multiply with the appropriate Born terms are given below.
\be
c^{VV} = 1+{\as\over2\pi}\CF\left[\,\tilde{\Gamma}-v{\xi\over2+\xi}\ln\left(
         {1+v\over1-v}\right)-{4\over v}\I{2}-{\xi\over v}\II{3}+
         {4\over v(2+\xi)}+{2-\xi\over v}\II{5}+\Delta_0^{VV}\,\right].
\label{cvv}
\ee
In this equation, the contribution from the virtual gluon loop is denoted as
\bea
\tilde{\Gamma} &=&
  \left[\,2-{1+v^2\over v}\ln\left({1+v\over1-v}\right)\right]\,
  \ln\left(\Frac{1}{4}\xi\right)+{1+v^2\over v}\left[\,
  \Li\left(-{2v\over1-v}\right)-\Li\left({2v\over1+v}\right)+
  \pi^2\,\right] \nn
  &&+3v\ln\left({1+v\over1-v}\right)-4.
\eea
Here, the $q\bar{q}g$ phase-space integrals are abbreviated by $\I{i}$,
and $\II{i}$ specify the results after the (soft) IR divergences have canceled.
The explicit analytical expressions for these phase-space integrals may be
found in \cite{tbp}. The additional component stemming from the anomalous
gluon bremsstrahlung is given by
\be
\Delta_0^{VV} = {8\over(2+\xi)v}\left[\,\real(\mu)(\I{1}+\I{4})+{2\over\xi}
  \left(|\mu|^2+|d|^2\right)(\I{1}-2\I{8})\,\right].
\ee
For the $AA$ contribution we find:
\be
c^{AA} = 1+{\as\over2\pi}\CF\left[\,\tilde{\Gamma}+2{\xi\over v}\ln\left(
         {1+v\over1-v}\right)+{\xi\over v^3}\I{1}-{4\over v}\I{2}-
         {\xi\over v}\II{3}+{2+\xi\over v^3}\I{4}-{2-\xi\over v}\II{5}
         +\Delta_0^{AA}\,\right],
\label{caa}
\ee
with the following anomalous part
\bea
\Delta_0^{AA} &=&
   {2\over v^3}\left[\,\real(\mu)\Big\{-(4-\xi)\I{1}+(2+\xi)\I{4}\Big\}
   +\left(|\mu|^2+|d|^2\right)\left\{\left({4\over\xi}+\xi-6\right)\I{1}+
   \xi\I{4}
   \right.\right.\nn&&\left.\left.
   -{4\over\xi}(2-\xi)\I{8}+{4\over\xi}\I{9}\right\}\,\right].
\eea
The remaining $V\!A$ and $AV$ parts are identical and only contribute to the
spin-dependent cross section. In the absence of anomalous couplings $(\mu=d=0)$, the cross section for longitudinally polarized quarks of helicity $\pm \half$
is given by
\be
\sigma\left(e^+ e^-\to\gamma,Z\to q(\lambda_\pm)\,\bar{q}\right) =
\Frac{1}{2}v(3-v^2)\sigma^{VV}c^{VV}+v^3\sigma^{AA}c^{AA}
\pm v^2\sigma^{V\!A}_S c_\pm^{VA}.
\ee
The multiplication factors $c^{ij}$ are expressed in terms of phase-space
integrals of type $\S{i}$:
\be
c^{V\!A,AV}_\pm =
1+{\as\over2\pi}\CF\left[\,\tilde{\Gamma}+{\xi\over v}\ln\left({1+v\over1-v}
\right)+\Delta^{V\!A,AV}_{\mu=d=0}\,\right],
\label{cva}
\ee
where
\bea
\Delta^{V\!A,AV}_{\mu=d=0} &=&
\Frac{1}{2}\Big[\,(4-\xi)\S{1}-(4-5\xi)\S{2}-2(4-3\xi)\S{4}-
\xi(1-\xi)(\SS{3}+\SS{5})+\xi(\S{6}-\S{7}) \nn
&& -2\S{8}+(2-\xi)\S{9}+(6-\xi)\S{10}-2\S{11}+
2(1-\xi)(2-\xi)\S{12}\,\Big].
\eea
The full analytic forms for the $S$ integrals are too lengthy to be exhibited
here. Most of them are compiled in Ref.~\cite{tbp}, except for the four
additional integrals:
\bea
S_{14} &=& {2\over\xi}-{2+\sxi\over2(2-\sxi)}-\ln(2-\sxi)+\Frac{1}{2}
           \ln\xi-\Frac{1}{2}, \\
S_{15} &=& \Frac{1}{32}\xi^3\Big[\Frac{1}{2}\ln\xi-\ln(2-\sxi)\Big]-
           \sxi\left(4+\Frac{1}{3}\xi+\Frac{1}{4}\xi^2\right)+\Frac{1}{8}
           \left(7-\half\xi\right)\xi+\Frac{1}{3}, \\
S_{16} &=& -{\xi^4(4-\xi)\over512(2-\sxi)^2}+
           \Frac{1}{16}\left(\Frac{3}{16}\xi-1\right)\ln(2-\sxi)
           -\Frac{3}{8}\xi\left(1+\Frac{3}{8}\xi\right) \nn
       & & +\Frac{1}{512}\xi^3\Big[4-\xi+(16-3\xi)\ln\xi\Big]
           +\Frac{1}{4}\left(\Frac{7}{3}-
           \Frac{1}{2}\xi+\Frac{1}{16}\xi^2\right)+\Frac{1}{24}, \\
S_{17} &=& \Frac{1}{32}\xi^2(6-\xi)\Big[-\Frac{1}{2}\ln\xi+\ln(2-\sxi)\Big]
           +\Frac{1}{128}(4-\xi){\xi^3\over(2-\sxi)^2} \nn
       & & +\Frac{1}{4}\xi^{3\over2}\left(\Frac{5}{3}-\Frac{1}{8}\xi\right)
           -\half\xi\left(1-\Frac{1}{64}\xi^2\right)+\Frac{1}{12}.
\eea

Including spins for the quark or the antiquark introduces additional
spin-flip terms in the $O(\as)$ $c$ factors given in Eqs.~(\ref{cvv}),
(\ref{caa}) and (\ref{cva}).
For longitudinal quark polarization we find
\be
c^{ij}_\pm =
\Frac{1}{2}\left[\,c^{ij}\pm{\as\over2\pi}\CF\Delta^{ij}_S\,\right].
\ee
The individual parity-parity combinations are
\bea
(2+\xi)v\,\Delta^{VV}_S &=&
 8\imag\left(\mu^*d\right)\Bigg[\,
-2\left(1-{2\over\xi}\right)\S{1}+\left(1-{4\over\xi}\right)\S{8}-\S{9}-
\S{10}+\S{11} \nn
&&\hskip2.4cm+2\left(1-{4\over\xi}\right)\S{13}+{4\over\xi}(\S{15}+\S{17})\,
\Bigg] \nn
&&+\imag(d)\Bigg[\,8(1-\xi)\S{1}-\Big\{8-3\xi(2-\xi)\Big\}\S{2}+
\Big\{8+\xi(2-3\xi)\Big\}\S{4} \nn
&&\hskip1.75cm-\xi(2+\xi)\S{6}-4\S{8}+4(1-\xi)\S{9}-4(3+\xi)\S{10}+4\S{11} \nn
&&\hskip1.75cm+\xi(2-\xi)\S{14}
\,\Bigg], \\[.5cm]
v^3 \Delta^{AA}_S &=&
4\imag\left(\mu^*d\right)\Bigg[\,
{2\over\xi}(1-\xi)(2-\xi)\S{1}+\left(5-{4\over\xi}\right)\S{8}-
(1-\xi)(\S{9}+\S{10})+\S{11} \nn
&&+2\left(3-{4\over\xi}\right)\S{13}-
2\left(1-{2\over\xi}\right)\S{15}-{4\over\xi}\S{16}-
2\left(1-{4\over\xi}\right)\S{17}\,\Bigg] \nn
&&+\imag(d)\Bigg[\,
2\xi\S{1}-(1-\xi)(4-\xi)(\S{2}-\S{4})-\xi(1-\xi)\S{6}-(2-\xi)\S{8} \nn
&&+(2-3\xi)\S{9}-(6-5\xi)\S{10}+(2+\xi)(\S{11}+2\S{13})+\xi(1-\xi)\S{14}
\Bigg],\\[.5cm]
v^2 \Delta^{V\!A,AV}_S &=&
\real(\mu)\Bigg[\,
-\xi(\S{1}+\S{7})-2\S{8}+(2-\xi)(\S{9}+\S{10})-2\S{11}\,\Bigg] \nn
&&\pm2\,i\imag(\mu)\Bigg[\,
(2-\xi)\S{1}-\xi\S{10}-2\S{13}\,\Bigg] \\
&&+\left(|\mu|^2+|d|^2\right)\Bigg[\,-(4-\xi)\S{1}-\xi\S{7}+4\S{8}+\xi\S{9}+
(4-\xi)\S{10}-4\S{11}\,\Bigg]. \nonumber
\eea
Using charge conjugation in the final state, one can readily obtain the
corresponding expressions for (longitudinal) antiquark polarization. In the
following, we denote the antiquark results by an additional bar, {\it i.e.\/}
$\Deltab{ij}_S$:
\bea
\Deltab{VV}_S &=& \Delta^{VV}_S, \\
\Deltab{AA}_S &=& \Delta^{AA}_S,
\eea
where the following identities hold
\bea
\Deltab{V\!A} &=& -\Delta^{V\!A}_S\ =\ \left(\Deltab{AV}_S\right)^*, \\
\hbox{with}\qquad \Deltab{AV}_S &=& -\Delta^{AV}_S.
\eea

Considering the above expressions, we can construct the following
combinations of polarization asymmetries of $t$ and $\tbar$,
\beq
\Delta\sigma^{(+)} = \Frac{1}{2} \Big[\,
\sigma(\uparrow ) - \sigma(\downarrow)
-
\overline{\sigma}(\uparrow) + \overline{\sigma}(\downarrow)
\,\Big],
\eeq
\beq
\Delta\sigma^{(-)} = \Frac{1}{2} \Big[\,
\sigma(\uparrow) - \sigma(\downarrow)
+
\overline{\sigma}(\uparrow) - \overline{\sigma}(\downarrow)
\,\Big],
\eeq
where $\sigma (\uparrow)$, $\overline{\sigma} (\uparrow)$
refer respectively
to the cross sections for top and antitop with positive helicity, and
$\sigma (\downarrow)$, $\overline{\sigma} (\downarrow)$
are the same quantities with negative helicity. Of these, $\Delta\sigma^{(+)}$
is CP even and $\Delta\sigma^{(-)}$ is CP odd. This is obvious from the fact
that under C, $\sigma$ and $\overline{\sigma}$ get interchanged, while under
P, the helicities of both $t$ and $\tbar$ get flipped. Consequently, $\sigma$
and $\Delta\sigma^{(+)}$, both nonzero in standard QCD, receive contributions
from combinations of anomalous couplings which are CP even, viz., Im$(\mu)^2 +
\vert d\vert ^2$ and Re$(\mu)$. On the other hand, $\Delta\sigma^{(-)}$
vanishes in standard QCD, and in the presence of anomalous couplings it
depends only on the CP-odd variables Im$(\mu^*d)$ and Im$(d)$. That
$\Delta\sigma^{(-)}$ depends on the imaginary parts rather than the real parts
of a combination of couplings follows from the fact that it is even under
naive time reversal T$_{\rm N}$, i.e., reversal of all momenta,
without change in helicities, and without interchange of initial and
final states (as would have been required by genuine time reversal). As a
consequence, it is odd under CPT$_{\rm N}$, and imaginary parts of couplings
have to appear in order to avoid conflict with the CPT theorem.

We concentrate on the CP-odd combination of top and antitop polarizations, 
since the CP-even combination would get contributions from higher-order QCD
as well any new interactions beyond the standard model. The CP-violating  
contribution cannot get nonzero contribution from QCD in any order, and
therefore a  nonzero chromoelectric moment would signal new physics beyond
the standard model.

Fig.~2a shows a three-dimensional plot of $\Dsigmam$  
as a function of Im$(\mu^*d)$ and Im$(d)$ at $\sqrt{s}=500$ GeV. Fig.~2b shows
a similar plot for $\sqrt{s}=1000$ GeV.

We use our expressions to obtain simultaneous 90\% confidence level (CL)
limits that could be obtained at a future linear collider with an integrated
luminosity of 50 fb$^{-1}$. We do this by equating the magnitude of the
difference between the values for a quantity with and without anomalous
couplings to 2.15 times the statistical error expected. Thus, the limiting
values of Im$(\mu^* d)$ and Im$(d)$ for an integrated
luminosity $L$ and a top detection efficiency of 
$\epsilon$ are obtained from
\beq
\epsilon\, L\,\left\vert\,\Dsigmam(\mu,d)-\Dsigmam_{\scriptscriptstyle SM}
\,\right\vert = 2.15 \,\sqrt{L\,\left\vert\,\sigma_{\scriptscriptstyle SM}
(\uparrow)+\overline{\sigma}_{\scriptscriptstyle SM}(\uparrow)\,\right\vert\:}.
\label{lim-}
\eeq
In the above expressions, the subscript ``SM" denotes the value expected in the
standard model, with $\mu=d=0$. We use $\epsilon = 0.1$
in our numerical estimates. For the running of the strong coupling, we choose
$\as^{(5)}(\MZ)=0.118$ (with $\MZ=91.178$~GeV) in the modified minimal
subtraction scheme and use the appropriate conditions to match for six active
flavours\footnote{It is common to indicate the number of active flavours as
superscript of $\as$. For practical purposes, one usually selects bottom
production as reference. Here, our choice for $\as^{(5)}(\MZ)$ translates to
$\as^{(6)}(M_t=172.1{\rm GeV})=0.10811$}.

Eq.~(\ref{lim-}) is used to obtain contours in the plane
of Im$(\mu^*d)$ and Im$(d)$,
shown in Figs.~3a and 3b for $\sqrt{s}=500$~GeV and 1000~GeV,
respectively. The contours are presented for
different $e^-$ longitudinal beam polarizations $P_-$.
In Fig.~3, the allowed regions are the bands lying
between the upper and lower straight lines.

A conclusion that can be drawn from Fig.~3 is that a large
left-handed polarization leads to increase in sensitivity.

For a fixed energy, there is only one CP-violating quantity which can be
measured, viz., $\Dsigmam$. This cannot give independent limits on the two 
quantities Im$(\mu^*d)$ and Im$(d)$. However, if a measurement of 
$\Dsigmam$ is made at two c.m.\ energies, a
relatively narrow allowed range can be obtained, allowing independent limits
to be placed on both Im$(\mu^*d)$ and Im$(d)$. This is demonstrated in Fig.~4. 
In fact, the improvement in the limit on Im$(\mu^*d)$ in going from
$\sqrt{s}=500$~GeV to $\sqrt{s}=1000$~GeV is considerable. The possible
limits are
\beq
-0.8 < {\rm Im} (\mu^* d) < 0.8,\
-11 < {\rm Im} (d) < 11.
\eeq
It is customary to use units of $e\,{\rm cm}^{-1}$ for the electric 
dipole moment.
In our case of the chromoelectric moment, we can express the above limits in
terms of an analogous unit, viz., $g_s{\rm cm}^{-1}$:
\beq
\vert {\rm Im} (\mu^*d) \vert< 9.2\times 10^{-17} g_s{\rm cm}^{-1},\
\vert {\rm Im} (d)\vert < 1.3\times 10^{-15} g_s{\rm cm}^{-1}.
\eeq
These limits may be compared with the limits obtainable from gluon jet energy
distribution in $e^+e^- \rightarrow t\tbar g$ \cite{rizzog}. While our proposal
for the CP-even case seems to fare worse,  for the CP-odd case, our proposal
can be competitive. It should however be emphasized that in the case of the
CP-odd couplings, we are proposing the measurement of a genuinely CP-violating
quantity, whereas the analysis in \cite{rizzog} is merely based on the energy
spectrum resulting from both CP-odd and CP-even couplings.
In case of $\Dsigmam$, the dependence on $e^-$ beam
polarization is rather mild.

We should also emphasize that we have taken the same integrated luminosity, viz.\
50~pb$^{-1}$, for c.m.\ energies of 500 GeV and 1000 GeV. It is usually assumed
that linear colliders operating at higher energies would simultaneously increase
their luminosities to off-set the drop in cross sections with energy. With a
higher integrated luminosity for 1000 GeV, our limits would be considerably
better than what we have obtained above with a somewhat conservative approach.

It is worthwhile noting that we have used a rather moderate value of
$\epsilon = 0.1$ for top detection and polarization analysis. A better efficiency
would lead to an improvement in the limits, as would a higher overall 
luminosity.

We have not considered the effect of initial-state radiation in this work. We
have also ignored possible effects of collinear gluon emission from one of the
decay products of $t$ or $\tbar$.  A complete analysis should indeed
incorporate these effects, as well as a study of $t$ and $\tbar$ decay
distributions which can be used to measure the polarizations.
However, we do not expect our conclusions
to change drastically when these effects are taken into account.

In summary, we have examined the capability of single
quark polarization in $e^+e^- \rightarrow t\tbar$ to measure or put limits on
the top chromoelectric dipole coupling. 
The CP-violating combination of top and
antitop polarizations is sensitive to the anomalous chromoelectric coupling,
and can yield a
limit of the order of 1 on a CP-odd combination of anomalous couplings.
\newpage
{\bf Acknowledgements.} We are both grateful for the kind hospitality and
stimulating scientific atmosphere we enjoyed at the Departament de F\'\i sica
Te\`orica during the principal stages of this work. This work has been
supported by the DGICYT under Grants Ns.\  PB95-1077 and SAB95-0175, as well
as by the TMR network ERBFMRXCT960090 of the European Union (S.D.R.).
M.M.T.\ acknowledges support by CICYT Grant AEN-96/1718 and the Max-Kade
Foundation, New York, NY.

\newpage
\thispagestyle{empty}
\centerline{\bf\Large Figure Captions}
\vskip1cm
\newcounter{fig}
\begin{list}{
   \bf Fig.~\arabic{fig}:\ }{
         \usecounter{fig}
         \labelwidth1.6cm
         \leftmargin2cm
         \labelsep0.4cm
         \itemsep0ex plus0.2ex
        }
\item Additional Feynman diagrams contributing to
      $\sigma\left(e^+e^-\to\gamma,Z\to t\bar{t}\right)$
      that account for anomalous gluon couplings at \oas.
      The large dots represent anomalous $t\bar{t}g$ insertions
      according to the effective action Eq.~(1).

\item Surface plots displaying the dependence of the polarization asymmetry
      $\Dsigmam$ on Im$(\mu^*d)$ and Im$(d)$ 
      with initial
      electron beam polarization $P_-=-1$ and c.m.\ energies {\bf(a)}
      $\sqrt{s}=500$~GeV, {\bf(b)} $\sqrt{s}=1000$~GeV.

\item Contour plots showing the allowed regions for
      $\Dsigmam$ with 90\% confidence level (integrated
      luminosity $L=50$~fb$^{-1}$ and top detection efficiency $\epsilon=0.1$).
      Representative c.m.\ energies are {\bf(a)} $\sqrt{s}=500$~GeV and
      {\bf(b)} $\sqrt{s}=1000$~GeV for various longitudinal electron
      polarizations.

\item Intersecting area resulting from
      two independent $\Dsigmam$ measurements at $\sqrt{s}=500$~GeV, 1000~GeV.
\end{list}
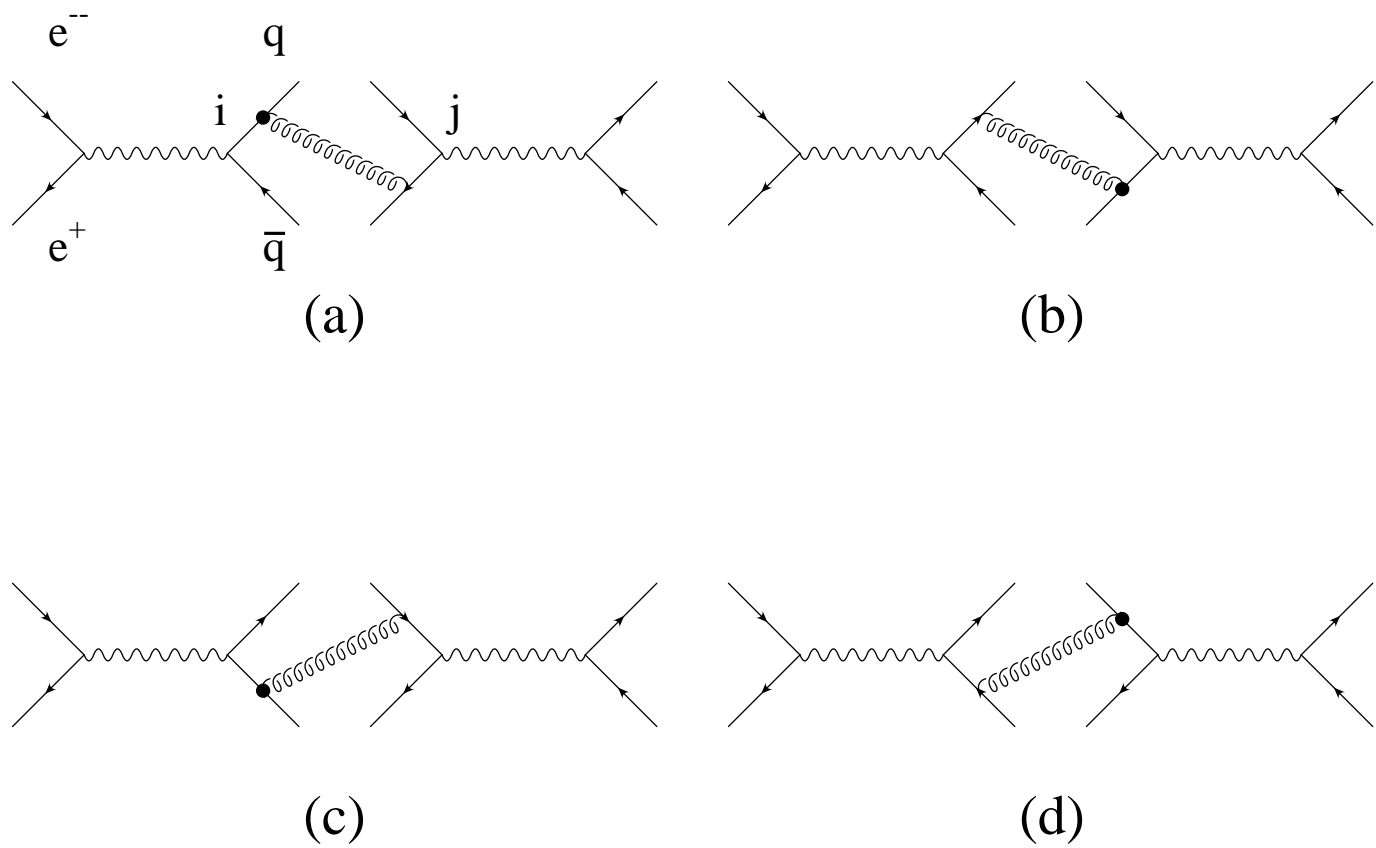
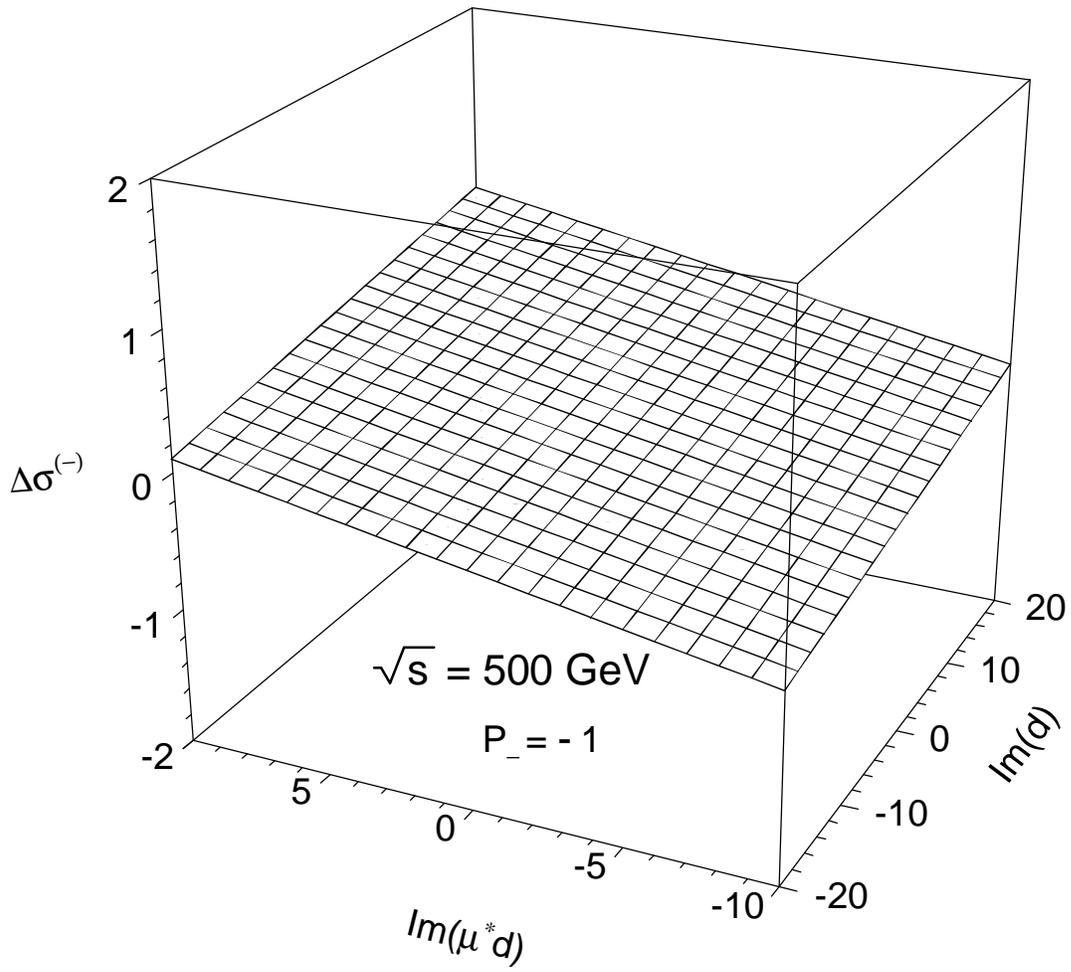
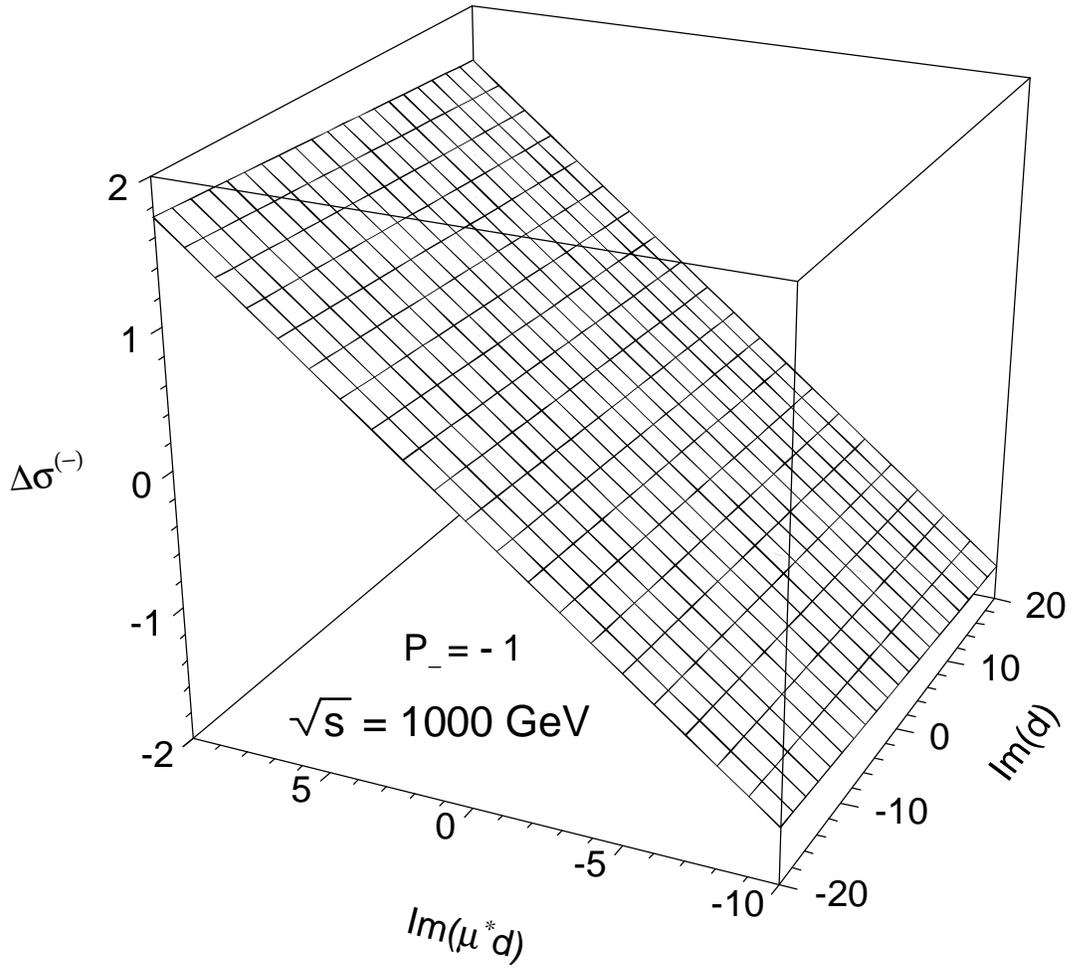
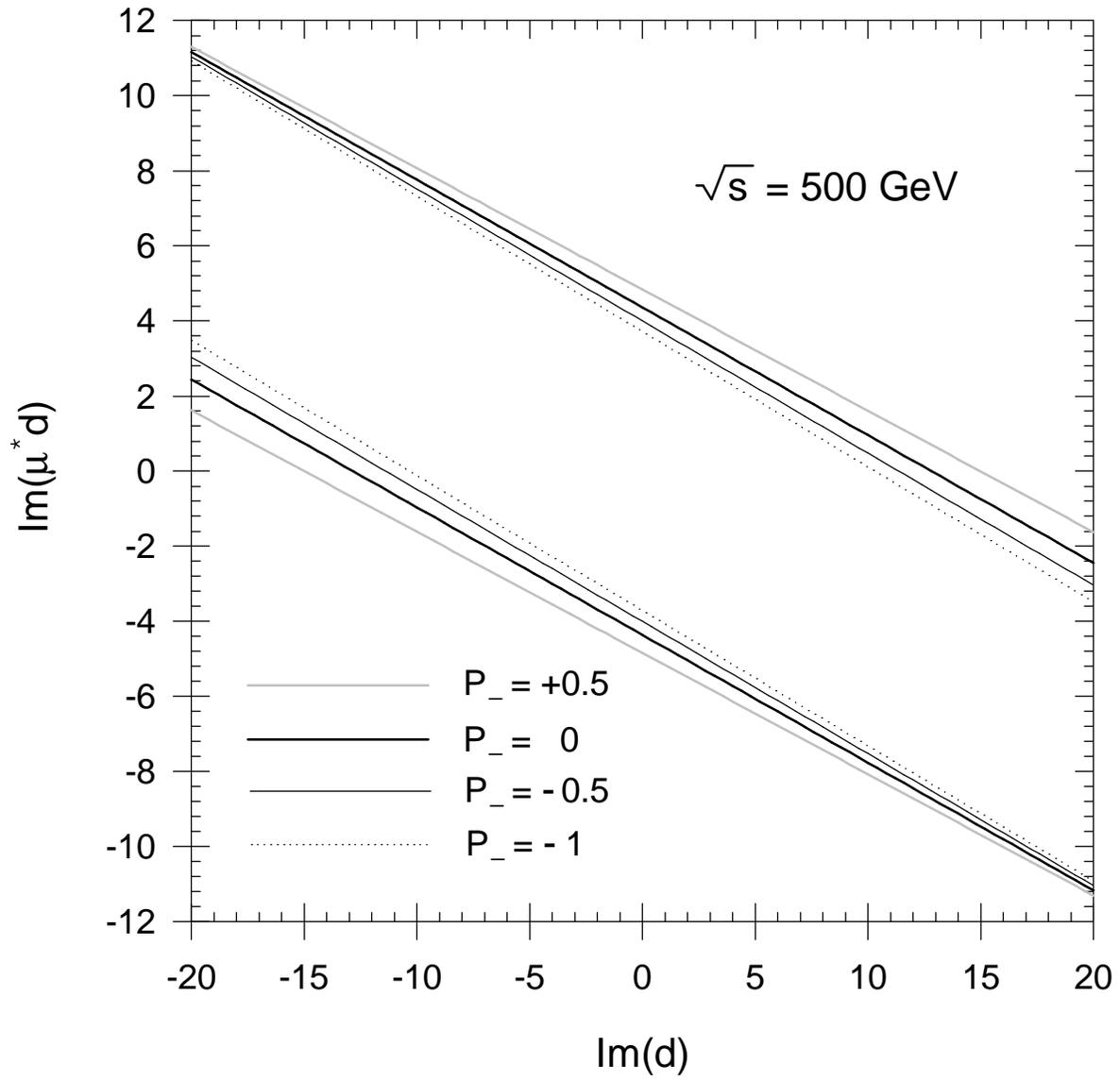
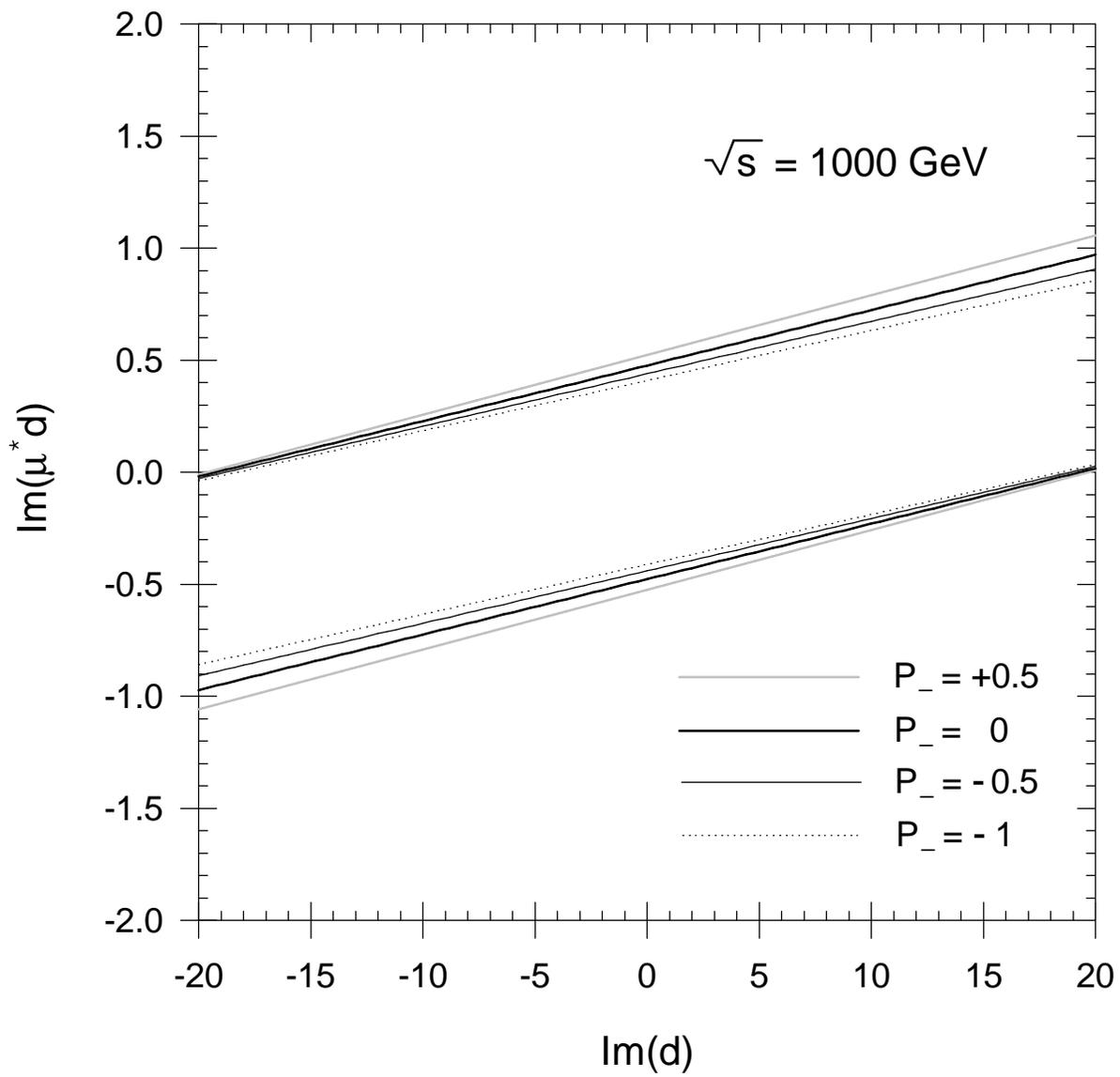
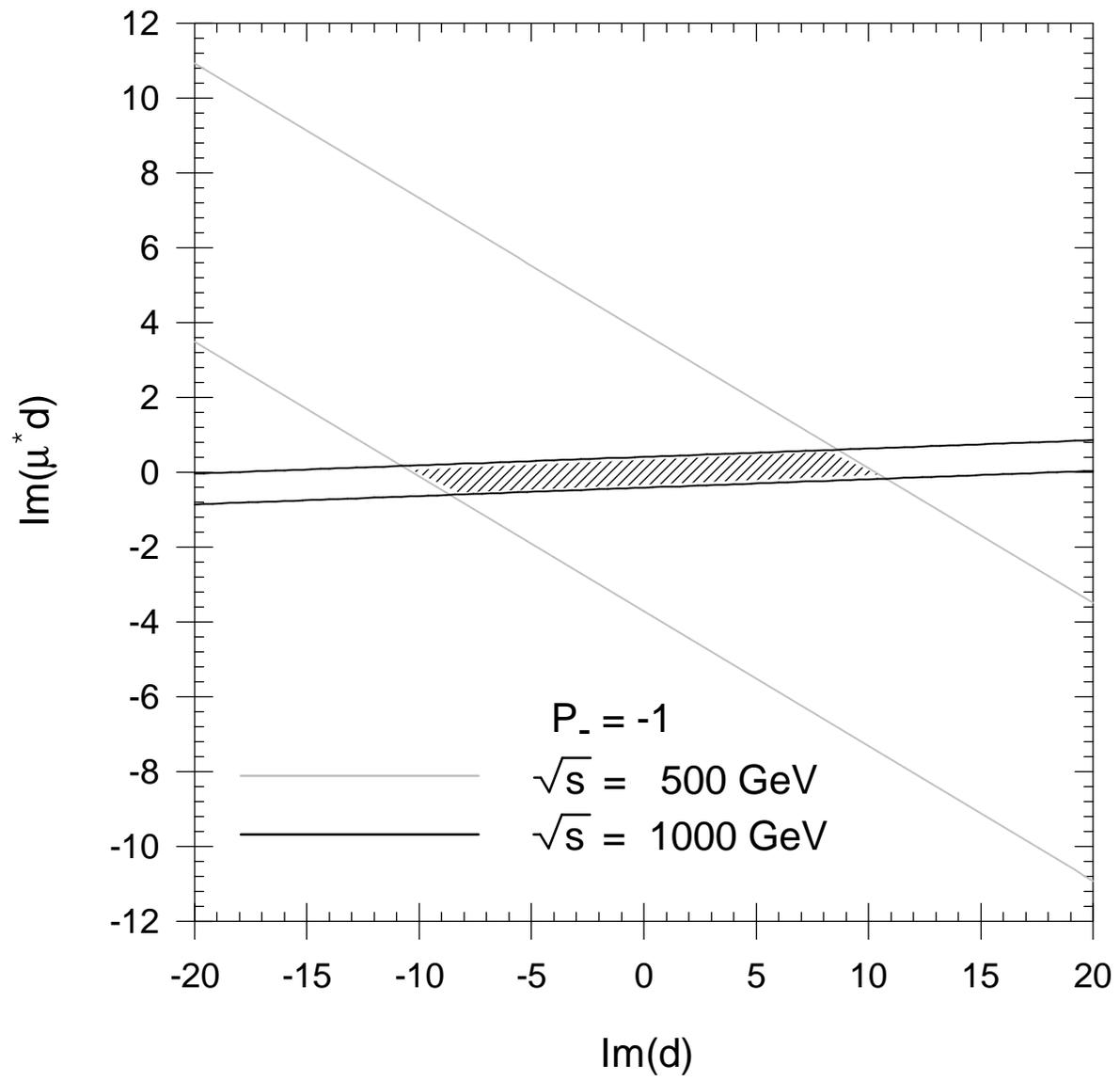
\end{document}